# Suppression of metal-to-insulator transition and stabilization of superconductivity by pressure in Re$_3$Ge$_7$

S. Huyan[1,2], E. Mun[3], H. Wang[4], T. J. Slade[1,2], Z. Li[1,2], J. Schmidt[1,2], R. A. Ribeiro[1,2], W. Xie[4], S. L. Bud'ko[1,2], P. C. Canfield[1,2]

[1] *Ames National Laboratory, US DOE, Iowa State University, Ames, Iowa 50011, USA*
[2] *Department of Physics and Astronomy, Iowa State University, Ames, Iowa 50011, USA*
[3] *Department of Physics, Simon Fraser University, Burnaby, BC V5A 1S6, Canada*
[4] *Department of Chemistry, Michigan State University, East Lansing, MI 48824, USA*

e-mail address: shuyan@iastate.edu
canfield@ameslab.gov

The effect of pressure on the low-temperature states of the Re$_3$Ge$_7$ is investigated by both electrical and Hall resistance and magnetization measurements. At ambient pressure, the temperature dependent resistance of Re$_3$Ge$_7$ behaves quasi-linearly from room temperature down to 60 K, then undergoes a two-step metal-to-insulator transitions (MIT) at temperatures $T_1$ = 59.4 K and $T_2$ = 58.7 K which may be related to a structural phase transition or occurrence of charge density wave ordering. Upon applying pressure, the two-step ($T_1$, $T_2$) MIT splits into three steps ($T_1$, $T_2$ and $T_3$) above 1 GPa, and all traces of MITs are fully suppressed by ~8 GPa. Subsequently, the onset of bulk superconductivity (SC) occurs between 10.8 and 12.2 GPa and persists to our highest pressure of 26.8 GPa. At 12.2 GPa the superconducting transition temperature, $T_c$, and upper critical field, $H_{c2}$ reach the maximum of $T_c$ (onset) ~5.9 K and $H_{c2}$ (1.8 K) ~ 14 kOe. Our results not only present the observation of SC under high pressure in Re$_3$Ge$_7$ but also delineate the interplay between SC and other competing electronic states by creating a $T$ - $p$ phase diagram for this potentially topologically nontrivial system Re$_3$Ge$_7$.

## INTRODUCTION

Superconductivity (SC) frequently arises in a proximity to competing phase transitions, wherein the competing phase is sufficiently suppressed, allowing the emergence of the superconducting state. These competing phases may include antiferromagnetic (AFM) transitions [1-3] and charge density wave (CDW) transitions [4], sometimes accompanied by a transition from a metallic to a nonmetallic electron state [5,6]. The conventional SC, characterized by an effective attractive interaction between electrons mediated through electron-phonon coupling, follows a similar pattern. The SC state can be induced by suppressing these competing states by adjusting parameters such as chemical substitution [7] and external pressure [8]. Understanding the interplay between these competing phases and superconductivity not only provides insight into the pairing mechanism of superconductivity, but also offers a viable method for investigating new superconductors.

Re$_3$Ge$_7$ is a material predicted to be a topological semimetal with a high symmetry point [9]. Recently, Rabus and Mun [10] grew single crystals and reported that Re$_3$Ge$_7$ undergoes a metal-to-insulator transition (MIT) at approximately 58 K. One report of temperature-dependent X-ray diffraction measurements show what appears to be an abrupt decrease of all of the unit cell parameters (and volume) upon cooling through the MIT temperature [11]. In contrast, in another report on polycrystalline samples, no clear abrupt change in lattice parameters and volume through MIT temperature was observed [12]. In addition, superconductivity was reported in Re$_3$Ge$_{7-x}$Ga$_x$ polycrystalline samples in which the Ga substitution completely suppressed the MIT and induced a superconducting dome [12]. Unfortunately, the Ga substitution simultaneously acted as hole doping, anisotropically changed the size of the unit cell parameters, decreased the unit cell volume and induced disorder. In order to systematically investigate the evolution of Re$_3$Ge$_7$ from a low temperature insulator to a low temperature superconductor we studied the physical properties of single-crystalline Re$_3$Ge$_7$ samples under high pressure.

Here we report a comprehensive high-pressure investigation on single crystals of Re$_3$Ge$_7$, which included electrical transport and magnetization measurements. Our findings indicate that the MIT features are suppressed monotonically by pressures up to 8 GPa. Around 2.4 GPa, the first trace of superconductivity was detected and between 10.8 and 12.2 GPa, bulk superconductivity is detected by both electrical resistance and magnetization measurements. In addition to the above general picture, we observed, at low temperature, a clear sign-change in Hall coefficient ($R_H$) at 2.4 GPa. Based on the aforementioned observations, a $T$-$p$ phase diagram is constructed, which highlights the emergence of SC in the pressure range associated with the complete suppression of the higher temperature MIT.

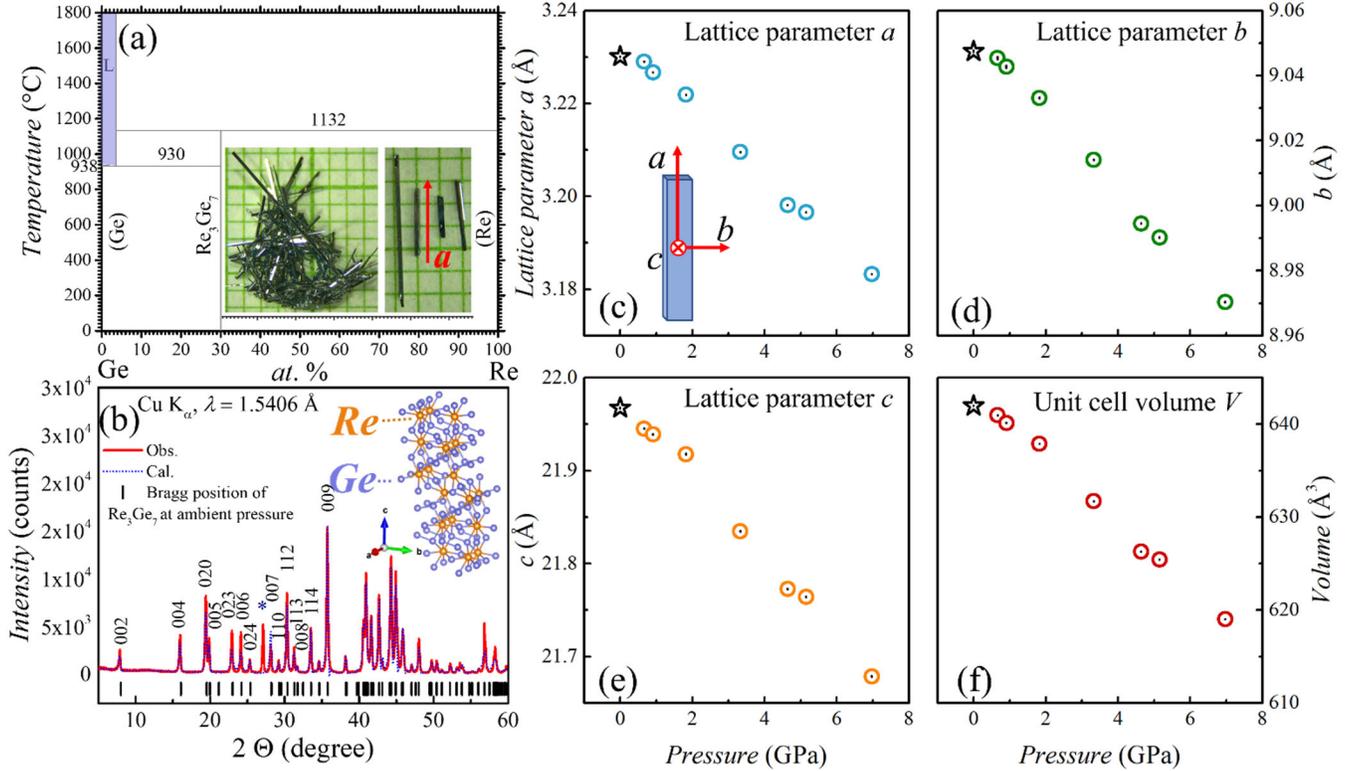

Fig. 1 (a) Ge-Re temperature-composition, binary phase diagram (ASM Diagram #901211) [13]. Inset shows the photograph of as-grown crystals on mm grid. The red arrow shows the *a*-axis of the crystal. (b) Powder X-ray diffraction patterns of $Re_3Ge_7$ sample. Inset shows the orthorhombic crystal structure of $Re_3Ge_7$. '*' in (b) denotes the unknow peak that does not belong to $Re_3Ge_7$ phase. (c)-(f) Pressure dependence of lattice parameters and unit cell volume of $Re_3Ge_7$ at room temperature up to 7 GPa. The lattice parameters and the volume were estimated from the refinement of the powder XRD pattern generated from 2D images obtained in the single crystal XRD measurements, shown in the appendix, Fig. A1. Black stars in (c)-(f) show the ambient pressure data (not in DAC). Inset in (c) shows crystallographic directions.

## Method

*Crystal growth and structure*

Given that the reported Re-Ge binary phase diagram [13] has a relatively limited (in temperature), essentially vertical liquidus line for the formation of $Re_3Ge_7$ (Fig. 1(a)), the controlled nucleation and growth of $Re_3Ge_7$, at first glance, appears to be problematic. As can be seen, $Re_3Ge_7$ is reported to decompose peritectically near 1130 °C and the nominally exposed liquidus line ends, at low temperatures, at the eutectic point with a reported temperature of 930 °C. Attempts to intersect the liquidus line between these temperatures (1130 - 930 °C) indeed prove frustrating given that slightly too much Re would lead to starting the growth from a two phase region (liquid and undissolved Re) giving rise to poorly controlled nucleation and slightly too little Re leads to a complete failure to grow $Re_3Ge_7$. Fortunately, given that we can use fritted crucible sets (Canfield Crucible Sets (CCS) [14,15], we can reliably and reproducibly grow $Re_3Ge_7$ in a two steps process.

First elemental Re (lumps 99.99 % purity) and Ge (pieces, 99.999%) were placed in a 2 ml crucible in a ratio that has Re much greater than 3% atomic; we used $Re_6Ge_{94}$. A frit and catch crucible [14,15,16] were placed on top of the growth crucible and sealed into a silica tube. This ampoule was then heated to 1200 °C, dwelling there for several days and then cooled to 1130 °C after which the ampoule was removed from the furnace and promptly decanted using a centrifuge [16]. On the growth side there was a lump of unconsumed Re and on the catch side there was now a $Re_xGe_{1-x}$ mixture that, when resealed and reheated to 1130 °C was ideally saturated and ready for growth of $Re_3Ge_7$. The decanted $Re_xGe_{1-x}$ material now becomes the growth crucible with a frit and catch crucible placed on top of it and the whole CCS sealed in a silica ampoule. The ampoule is then heated to 1180 °C, dwells there for up to a day, cooled to 1140 °C and then cooled over 100 hours to 945 °C and then again, decanted. The crystals generally grow with a needle-like or ribbon-like morphology, with the longest dimension along the crystallographic *a* axis, as shown in the inset of Fig. 1.

The use of a fritted crucible set allows for the ready creation of saturated liquid at given temperature (the decanting temperature in the first step). The clean re-use of this decanted liquid allows for growth of crystals even when precise information about the location of the liquidus line is absent. This re-use of decanted liquid is also the key step in fractionation of a growth as is described in detail in reference [17].

The crystal structure was determined by powder X-ray diffraction (XRD) measurements in a Rigaku Miniflex diffractometer using Cu $K_\alpha$ (X-ray), shown in Fig. 1(b). The detected peaks in the powder XRD spectra agree well with previous report [10,18]. The strongest diffraction peak of the elemental Ge is also detected which comes from the Ge flux attached on the crystal surface that was not fully removed during the decanting process. The calculated lattice parameters according to the diffraction peaks give $a$ = 3.23(4) Å, $b$ = 9.04(5) Å, $c$ = 21.96(5) Å, which is consistent with the literature values [18]. The crystal structure of Re$_3$Ge$_7$ is shown in the inset of Fig. 1(b). The orientation of the crystal was determined by Laue backscattering and is depicted in the inset of Fig. 1(c). To investigate the pressure effects on the crystal structure, high-pressure single crystal XRD experiment was conducted on the single crystal of Re$_3$Ge$_7$ with the in-plane dimensions of 0.088 × 0.070 mm$^2$ up to 6.4 GPa at room temperature. Prior to the high pressure experiment, the sample was mounted on a nylon loop with paratone oil and measured at ambient pressure to confirm its structure. The sample was loaded in the Diacell One20DAC manufactured by Almax-easyLab [19] with 500 μm culet-size extra aperture anvils. The 250 μm thick stainless-steel gasket was preindented to ~110 μm and then a 210 μm diameter hole was drilled in the center using an electronic discharge machine drilling system. An methanol: ethanol = 4:1 mixture was used as the pressure transmitting medium while the pressure in the cell was determined by the $R_1$ fluorescence line of ruby [20]. The single crystal XRD experiments were performed using a Rigaku XtalLAB Synergy, Dualflex, Hypix single crystal X-ray diffractometer on Mo $K_\alpha$ radiation ($\lambda$ = 0.71073 Å, micro-focus sealed X-ray tube, 50 kV, 1 mA) at 300.0(1) K. The total number of runs and images was based on strategy calculation from the program CrysAlisPro (Rigaku OD). Data reduction was performed with correction for Lorentz polarization. For the ambient pressure dataset, numerical absorption correction based on gaussian integration over a multifaceted crystal model Empirical absorption correction using spherical harmonics, implemented in SCALE3 ABSPACK scaling algorithm. The crystal structure was solved and refined using the Bruker SHELXTL Software Package [21,22]. At 300 K and ambient pressure, Re$_3$Ge$_7$ was confirmed to crystallize in the orthorhombic structure (space group, *Cmcm*, #63). The powder XRD patterns were generated from 2D images obtained in the single crystal XRD measurements (shown in the appendix, Fig. A1). From the result, no new diffraction peaks were observed at pressure up to 6.4 GPa, suggesting that the sample maintains a consistent orthorhombic structure (*Cmcm*, #63) over the entire pressure range. On the other hand, as the pressure increased, the unit cell of Re$_3$Ge$_7$ began to compress, leading to a close-to-linear trend of decreasing in each of the three separate lattice parameters and the unit cell volume, see Fig. 1(c)-(f), indicating that the sample does not undergo any dramatic or conspicuous structural phase up to 6.4 GPa. It is important to note that whereas for Ga substitution [12] only the *c*-axis decreased, for applied hydrostatic pressure all axes decreased.

### High pressure measurements: resistance and magnetization

The electrical resistance measurements with current applied within the *ab* plane and magnetic field parallel to *c*-axis were performed in a Quantum Design Physical Property Measurement System (PPMS). The crystal was cleaved and cut into a thinner regular shape for the high pressure electrical measurements so that the most of excess Ge impurity could be avoided. A standard, linear four-probe method was used for measurement in piston-cylinder cell (PCC, Sample #1). An approximately standard four terminal configuration for Hall measurement was used in diamond anvil cell (DAC, Sample #2) (Bjscistar, [23]). We applied 90 kOe and -90 kOe magnetic field to get the precise magnetoresistance by ($R$ (90 kOe, $T$) + $R$ (-90 kOe, $T$)) / 2, and Hall resistance by ($R$ (90 kOe, $T$) – $R$ (-90 kOe, $T$)) / 2, respectively. To apply pressures up to ∼ 2.27 GPa, a PCC was used. A 4:6 mixture of light mineral oil: n-pentane, which solidifies at ~3.6 GPa at room temperature [24], was used as pressure transmitting medium. High purity Pb was used to measure the pressure at low temperature. To apply higher pressure, up to ∼ 26.4 GPa, a diamond anvil cell (DAC) that fits into a Quantum Design Physical Property Measurement System (PPMS) was used with 500 μm culet-size standard-cut standard cut-type Ia diamonds. The sample was cut and polished into an 80×40×15 μm thin flake and loaded together with a tiny piece of ruby ball into the 250 μm thick, apertured, stainless-steel gasket covered by cubic-BN. The sample chamber was about 150 μm in diameter. Platinum foil was used as the electrodes to connect to the sample. The Nujol mineral oil was used as pressure transmitting medium (PTM), since: 1) fluid medium could still maintain a quasi-hydrostatic pressure environment with small pressure gradient below a liquid/glass transition which occurs at ∼ 15 GPa [25-28]; 2) the use of fluid medium avoids the direct contact between the sample and diamond culet which will lead to a further contribution of uniaxial pressure component. Pressure was determined by $R_1$ line of the ruby florescent spectra. [20] It is noteworthy that in the PCC run, as shown in Fig. 2(a), a tiny resistance anomaly at ∼ 3.5 K could be observed, which was not observed in DAC run. We consider this to be extrinsic but have not identified its possible origin.

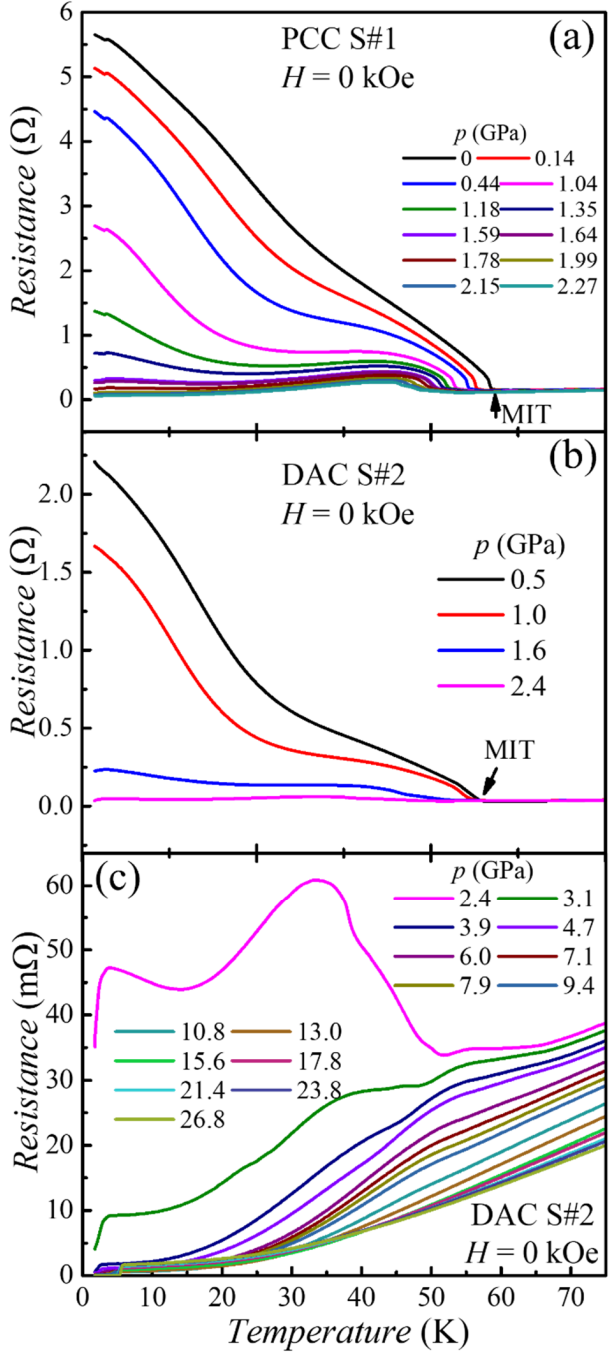

Fig. 2 **Resistance as a function of temperature under high pressure**. Resistance data at zero magnetic field are shown (a) for Piston Cylinder Cell (PCC); and (b) for Diamond Anvil Cell (DAC) 0.5 – 2.4 GPa and (c) DAC for 2.4 - 26.8 GPa.

The DC magnetization measurements under high pressure were performed in a Quantum Design Magnetic Property Measurement System (MPMS) at temperatures down to 1.8 K. The DAC (easyLab® Mcell Ultra [29]) with a pair of 500-μm-diameter culet-sized diamond anvils was used. The apertured Tungsten gasket with 300-μm-diameter hole was used to lock the pressure. The Nujol mineral oil, same as the DAC for electrical transport measurement, was used to keep the consistency of pressure environments in two sets of experiments. The applied pressure was measured by the fluorescence line of ruby ball [20]. The background signal of the DAC without sample was measured under 3.4 GPa in a 0.2 kOe applied field. Then the DAC was opened and re-closed after loading the sample with a dimension of 150 μm ×150 μm ×20 μm. The exact same measurements as previous background measurements were then performed at various pressures. The magnetization of the sample was analyzed by first performing a point by point subtraction of long-scan response with/without the sample, and then a dipole fitting of the subtracted long-scan response curve [30].

## Results and discussion

### Electrical resistance measurements

The data for the electrical resistance of $Re_3Ge_7$ under high pressure are summarized in Fig. 2. As shown in Figs. 2(a), and 2(b), for pressures $p$ < 2.4 GPa, the PCC and DAC data exhibit similar temperature-dependent resistance, $R(T)$, evolution at low temperatures. In general, as pressure increases, the MIT temperature decreases from ~ 58.5 K at ambient pressure to ~ 51 K at 2.4 GPa, the insulating behavior below the MIT temperature is progressively suppressed, and, for $p$ = 2.4 GPa, the material returns to metallic state at low temperatures. An increasingly clear resistance decrease is observed at low temperatures at and above 1.6 GPa and becoming clearer by 2.4 GPa. Since for $p$ ≥ 3.9 GPa, as depicted in Figs. 2(c) and 3(a), a clear resistance decrease could be observed, this downturn appears to be the onset of filamentary superconductivity. Starting at 10.8 GPa and continuing up to 26.8 GPa, pressure-induced SC with zero resistance becomes observable (see Fig. 3(a)).

### Superconducting properties

Figure 3 illustrates the superconducting behavior of $Re_3Ge_7$ under high pressure. The onset of the superconducting transition temperature, $T_c^{onset}$, shown in Fig. 3(a), is defined as the temperature at which the line of the highest slope of the resistance curve during the transition intersects with the straight-line fit of the normal state above the transition. $T_c^{onset}$, which is ~2.6 K at 3.9 GPa, increases to ~5.9 K at 13 GPa, and shows minor change up to 26.8 GPa. (It should be noted that we can identify $T_c^{onset}$ for even lower pressures in Fig. 2c, where for 2.4 GPa and 3.1 GPa we can infer 3.5 K and 3.3 K respectively). For pressures below 10.8 GPa, the superconducting transition is relatively broad and does not reach zero resistance, whereas for pressures at and above 10.8 GPa, the superconducting transition width becomes narrow, and zero resistance is observed. The magnetic susceptibility, Fig. 3(b), measure for $p$ = 12.2, 18.2, and 24.1 GPa, as a function of temperature at 0.2 kOe, shows a clear, unsaturating, diamagnetic signal down to 2 K, below $T_c$. The increase in pressure from 12.2 GPa to 24.1 GPa results in an increase in the estimated shielding volume fraction (where the demagnetization factor, N, is estimated to be 0.833 for a sample with dimensions of 150 μm ×150 μm × 20 μm and a magnetic field parallel to $c$-axis [31]), increasing from approximately 45% to about 80% at 2 K. As the diamagnetic transition does not show saturation, it is expected that the shielding volume fraction could approach closely 100% at high pressure or at lower temperature.

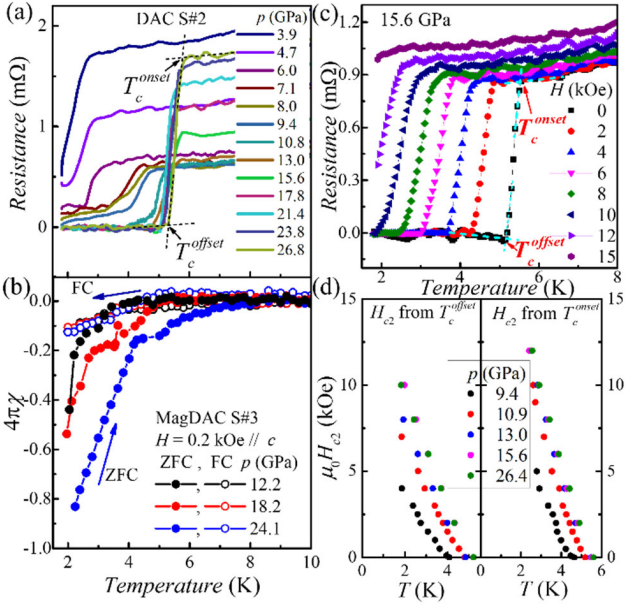

Fig. 3 **Superconductivity properties under pressure** (a) $R(T)$ in the low temperature range at various pressure from 3.9 GPa to 26.8 GPa, showing the superconducting transitions. (b) Magnetic susceptibility as a function temperature above 12.2 GPa at low temperature, showing diamagnetic transitions. Solid and hollow circles are zero field cool (ZFC) and field cool (FC) data, respectively. (c) $R(T)$ for $p$ = 15.6 GPa at different magnetic fields. The criteria of $T_c^{onset}$ and $T_c^{offset}$ in (a) and (c) are the intersections of two extended lines along the $R(T)$ curves above and below the transition temperatures as shown in the figure. (d) Upper critical fields as a function of temperature determined from $T_c^{offset}$ (left panel) and $T_c^{onset}$ (right panel) at different pressures.

To determine the upper critical field $\mu_0 H_{c2}$, we measured the temperature dependence of the resistance at various pressures under different magnetic fields, $p$ = 15.6 GPa is shown in Fig. 3(c) as an example. We can see that $T_c$'s are clearly suppressed by magnetic fields with only a slight broadening of the transition width. Zero resistance could not be observed above 12 kOe down to 1.8 K. The upper critical field ($H_{c2}$) as a function temperature with the $H_{c2}(T)$ values obtained from the offset and onset criteria are plotted in Fig. 3(d). For the highest pressures, we can estimate $H_{c2}(1.8$ K) from the Ginzburg-Landau model to be for onset and offset criteria, respectively. Further analysis of the $H_{c2}(T)$ data is presented in the appendix, Fig. A2.

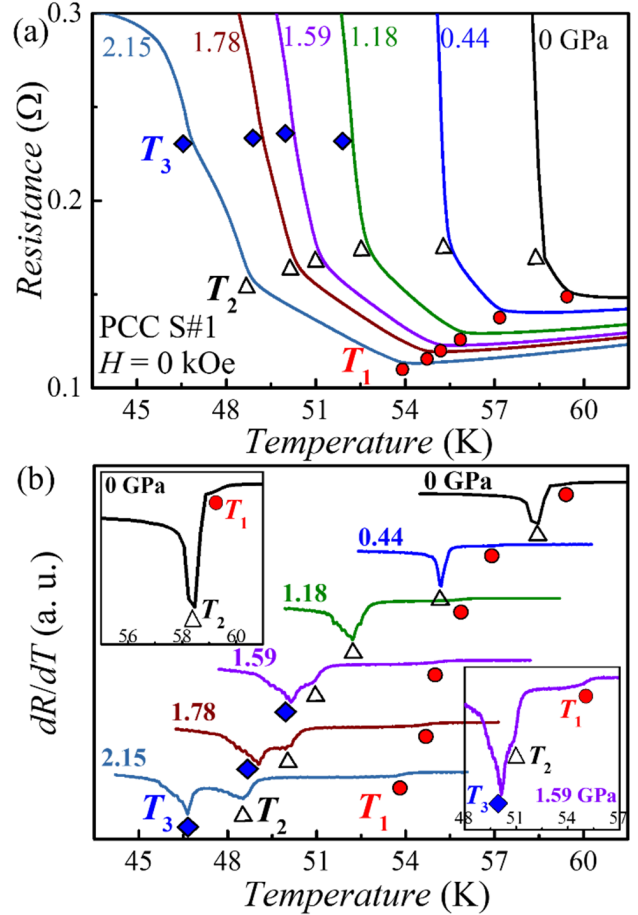

Fig. 4. **Determination of multiple transition temperatures from PCC run.** (a) $R(T)$ for selected pressures. (b) 1st temperature derivative resistance as a function of the temperature, $dR/dT$. Red filled circle, open triangle, and blue filled diamond in both (a) and (b) are used to show the locations of the transitions, $T_1$, $T_2$ and $T_3$. $T_1$, $T_2$ and $T_3$ are inferred from (b) by the criterions of: $T_1$ of the lower step edge of $dR/dT$ curve, and $T_2$&$T_3$ of the local minimum of $dR/dT$ curves, respectively. Insets in (b) identify the characteristic temperatures $T_1$, $T_2$ and $T_3$ at ambient pressure and 1.59 GPa.

## Determination of MIT temperature from resistance and Hall data

It is clear from the electrical resistance measurement data under pressure that, at a broad level, Re$_3$Ge$_7$ goes into a superconducting state upon the suppression of its MIT. Now, we investigate the evolution of MIT under pressure in greater detail. At ambient pressure, according to the $R(T)$ as well as *dR/dT* anomalies (demonstrates in Fig. 4(a) and the left inset of Fig. 4(b), $T_1$ and $T_2$ correspond to two phase transitions occurring around 59.4 K and 58.7 K, respectively. Both $T_1$ and $T_2$ decrease with increasing pressure until 1.59 GPa (as shown in the right inset of Fig. 4(b)), at which point the $T_2$ peak separates into two peaks, denoted $T_2$ and $T_3$ (Figs. 4(a), (b)). All three transitions continue to decrease with increasing pressure, until it becomes difficult to discern the transitions from the resistance data above 3.9 GPa (Figs. 5(a), (b)). Whereas at ambient pressure it could be tempting to dismiss the higher transition temperature as a slight broadening or artifact of some sort, the fact that pressure very clearly separates these features (and leads to the evolution of a new one) supports the idea that there are indeed multiple transitions associated with the ambient and low pressure MIT. This is further supported by the agreement between the PCC and low pressure DAC data.

We thus believe that the ambient and low pressure, two-step-MITs feature is intrinsic for the Re$_3$Ge$_7$ sample. Under high pressure, the transitions change from a fully gapped state to one with a cascade of partial gapping, similar to e.g. NbSe$_3$ [32], which has two partial gappings of the Fermi surface and still remains metallic down to lowest temperatures.

Under high pressure, the Hall coefficient ($R_H$) as a function of temperature was measured in order to further comprehend the suppression of the MIT, which typically reflects a significant change in Fermi surface topology. The $R_H$ curves are summarized in Fig. 5(c). A pronounced upward transition of $R_H$ is observed at around $T_1$ for low pressures, indicating a decrease in p-type carrier density, consistent with the observation of insulator-like resistance behavior below $T_1$. The dominant type of carrier is p-type. At 1.6 GPa, a pressure-induced reversal of the $R_H$ curve below $T_1$ from an upward to a downward transition was observed, indicating the change of carriers from p-type to n-type. This suggests that the nature of the excited carriers (hole versus electron) is delicately balanced in this material. The $T_1$ evolution could be traced up to 7.1 GPa with the clear $R_H$ transition visible near ~ 10 K. Uniquely identifying other transition temperatures in the Hall data is more difficult. We did not try to track $T_2$ or $T_3$ based on Hall data (larger scale Hall data is shown in Fig. A5).

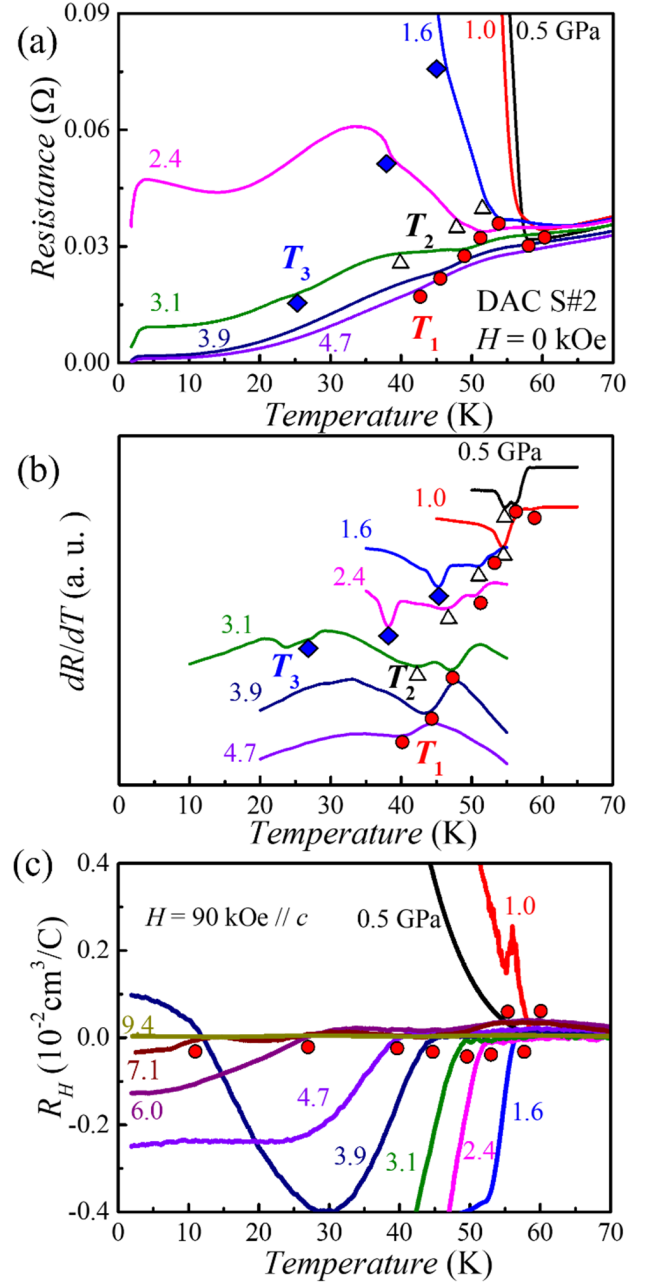

Fig. 5. **Determination of multiple transition temperatures from DAC transport data.** (a) Resistance as a function of the temperature, $R(T)$. (b) 1st temperature derivative resistance as a function of the temperature, *dR/dT*. (c) Hall coefficients as a function of the temperature, $R_H(T)$, with magnetic field, ±90 kOe, parallel to *c*-axis. Red filled circle, open triangle, and blue filled diamond in (a), (b), and (c) are used to show the locations of the transitions, $T_1$, $T_2$ and $T_3$. $T_1$, $T_2$ and $T_3$ are estimated from (b) by the criterions of: $T_1$ of the lower step edge of *dR/dT* curve, and $T_2$&$T_3$ of the local minimum of *dR/dT* curves, respectively.

## Phase diagram and discussion

Figure 6(a) presents a comprehensive $T$–$p$ phase diagram as well as the pressure dependence of the residual resistance ratio ($R_{300K}/R_{6K}$), and low temperature Hall coefficient ($R_H$). By increasing the pressure, the multiple transitions are progressively suppressed, and was estimated that the insulating state is suppressed by ~ 2.6 GPa and the density-wave-like states are suppressed by ~ 8 GPa. The suppression of the MIT is also clearly seen in the pressure dependent $R_{300K}/R_{6K}$ and $R_H(1.8K)$ data presented in Fig. 6(b). Both data sets indicate that, as pressures increased through 2.6 GPa, the sample becomes metal-like. Indeed, for $p \sim 10$ GPa $R_{300K}/R_{6K}$ is higher than 200, suggesting that $Re_3Ge_7$ becomes a very good metal. By 10.8 GPa there is a complete resistive transition to zero and for $p >12.2$ GPa there is a clear diamagnetic signal below $T_c$. $T_c$ onset (offset) values remain relatively constant at ~ 5.9 K (5.2 K) respectively for ~15 < $p$ < ~26 GPa.

The above results provide compelling evidence of a connection between superconductivity and the suppression of the transitions, that at ambient pressure resulted in a MIT, to zero Kelvin. Utilizing synchrotron X-ray diffraction, Verchenko et al. reported a clear isostructural transition below the MIT temperature, where all lattice parameters exhibited a sharp decrease, that was interpreted as a second-order phase transition [11]. Whereas it seems unlikely that such a discontinuous change in all three lattice parameters would be associated with a second order phase transition, our data are in clear disagreement with the sign of reported changes. If indeed all three lattice parameters decrease with pressure, then we would anticipate an increase of the MIT transitions with pressure, not a strong decrease. On the other hand, Cui et al. reported that no structural phase transition is observed through the MIT temperature via systematic low temperature powder XRD measurements, while, their DFT calculation suggested that MIT might be linked to Fermi surface nesting, potentially leading to the formation of charge density wave (CDW) ordering [12]. It is clear that further experimental (structural) and theoretical studies are needed to understand the origin of the MIT transition(s) in the $Re_3Ge_7$.

With the $T$ - $p$ phase diagram in mind, we can also return to the effects of pressure and Ga doping on the room temperature lattice parameters. Whereas the effect of Ga substitution and applied pressure are qualitatively similar, i.e., suppression of the MIT and onset of a SC phase, the effects of Ga and pressure on the lattice parameters are markedly different. Whereas the change in unit cell volume with Ga substitution needed to induce superconductivity is negligible (0.04%), for applied pressure a change in volume of roughly 3.5% is needed to induce superconductivity. These results suggest that whether it be by changes in band filling (via Ga doping) or via changes in lattice parameters (applied pressure) the suppression of the MIT is the necessary step to stabilizing superconductivity in $Re_3Ge_7$.

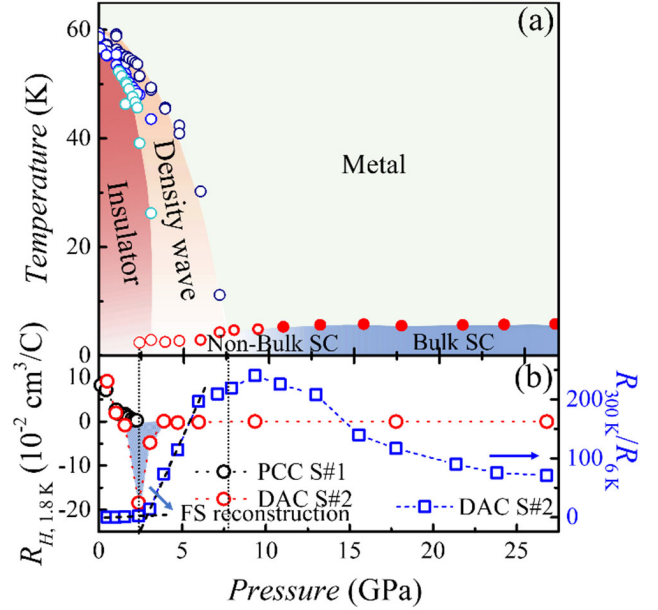

Fig. 6 **Schematic phase diagram**. (a) Temperature - pressure phase diagram of $Re_3Ge_7$. The red open circle and the red filled circle represent the non-bulk superconductivity (non-bulk SC) and bulk superconductivity (bulk SC), which is distinguished by if observing the zero resistance below $T_c$ onset, shown in Fig. 3(a). (b) Hall coefficient at 1.8 K (circles, left Y-axis) and residual resistance ratio ($R_{300\,K}/R_{6\,K}$, squares, right Y-axis) as a function of the pressure. The $R_{H,1.8\,K}$ data was estimated from the $R(H)$ data shown in the appendix, Fig. A6.

## Conclusions

In conclusion, the electrical transport properties and magnetic susceptibility of single crystal $Re_3Ge_7$ are investigated under high pressure. Upon applying pressure, the two-step MIT ($T_1$, $T_2$) evolves into a three-step MIT ($T_1$, $T_2$ and $T_3$) above 1 GPa. The multiple transition temperatures are suppressed by increasing the pressure, with the insulating state being suppressed by $p \sim 2.6$ GPa and with all transition temperatures being reduced to zero before reaching 8 GPa. SC reveals its bulk character at and above 10.8 GPa, and by 12.2 GPa the superconducting $T_c$ and $H_{c2}$(onset) become roughly pressure independent at values of 5.9 K and 18 kOe, respectively, up through our highest applied pressure of ~27 GPa.

## Acknowledgements

Work at Ames National Laboratory is supported by the US DOE, Basic Sciences, Material Science and Engineering Division under contract no. DE-AC02-07CH11358. T.J.S. was partially supported by the Center for Advancement of Topological Semimetals (CATS), an Energy Frontier Research Center funded by the U.S. Department of Energy Office of Science, Office of Basic Energy Sciences, through Ames National Laboratory. E. Mun is supported by the Canada Research Chairs program, the Natural Science and Engineering Research Council of Canada, and the Canadian Foundation for Innovation. H. Wang and W. Xie are supported by US DOE under contract no. DE-SC0023648.

## Appendix

The powder XRD pattern at 300 K under pressure generated from 2D images obtained in the single crystal XRD measurements, shown in Fig. A1, indicates that no new extra diffraction peak emerges under pressure, indicating that sample maintains the orthorhombic structure (*Cmcm*, #63) over the whole pressure range up to 6.4 GPa. With this said, it should be noted that starting at and above 1.2 GPa, there is a conspicuous and sharp increase in the intensity of the [004] peak at approximately 8°. On the other hand, an unexpected shift towards higher angles in the diffraction peak located around 21°. The reason for these phenomena remains unclear and underscores the need to have temperature dependent synchrotron diffraction data taken under pressure.

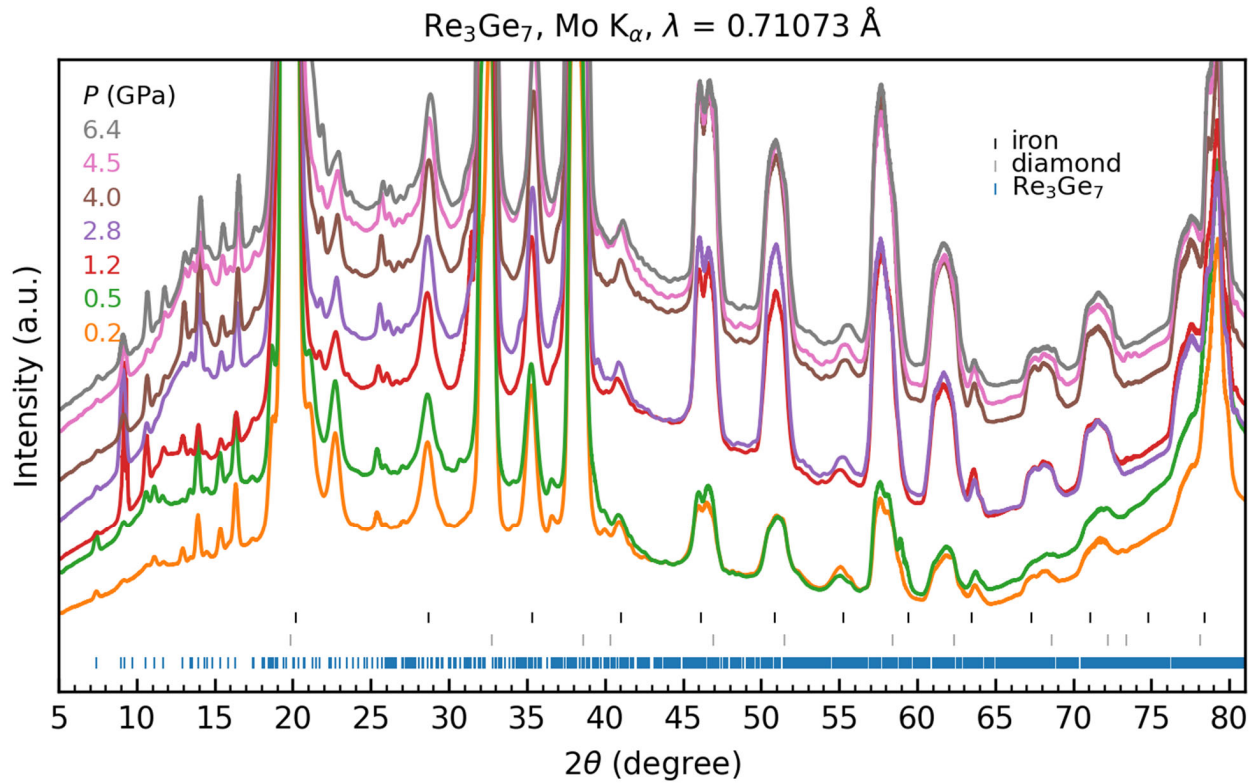

Fig. A1 Pressure dependent powder XRD pattern generated from 2D images obtained in the single crystal XRD measurement at 300 K.

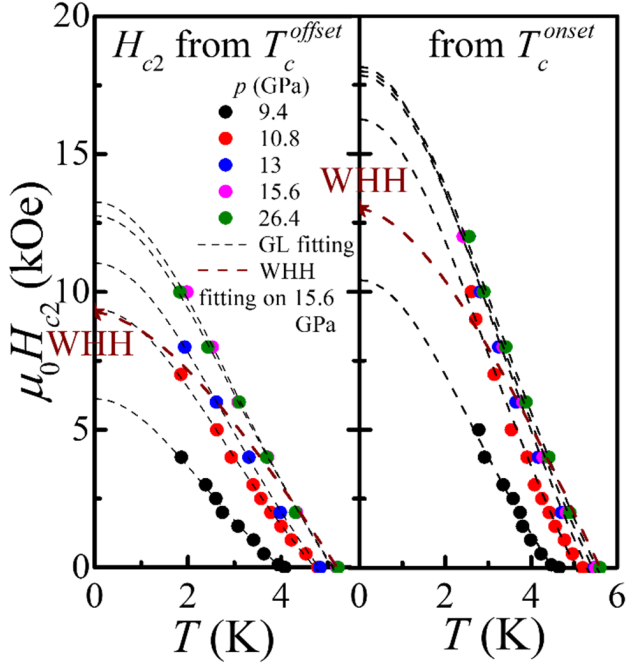

Fig. A2 Fitted curves by Ginzburg- Landau (G-L) model and Werthamer-Helfand-Hohenberg (WHH) model on the $H_{c2}$ data. G-L fittings were done on all data. WHH fitting was done on 15.6 GPa data only to show the difference between G-L and WHH fitting results.

$H_{c2}(0)$ is estimated by the Ginzburg-Landau (G-L) model, shown in Fig. A2. G-L formula is given as,

$$H_{c2}(T) = H_{c2}(0) \times \left[\frac{(1-t^2)}{(1+t^2)}\right] \quad (1)$$

where $t = T/T_c$ is the reduced temperature and $H_{c2}(0)$ is the upper critical field at zero temperature. Again, taking the 15.6 GPa data as example, denoting the $T_c = T_c^{onset}$, G-L fitting yields $H_{c2}(0)$ at 15.6 GPa around 18 kOe, which is within the Pauli paramagnetic limit in the case of weak coupling given by $1.84\ T_c \approx 104$ kOe. The G-L coherence length is 13.5 nm according to the formula.

$$\xi_{ab}(0) = \sqrt{\frac{h}{2e}\frac{1}{2\pi H_{c2,\perp}(0)}} \quad (2)$$

where $\frac{h}{2e} = \Phi_0$ is magnetic quantum flux constant. We also plot the universal curve for a clean spin singlet SC by the Werthamer-Helfand-Hohenberg (WHH) model [36,37] at 15.6 GPa with the orbital-limited upper critical field as

$$H_{c2}^{orb}(0) = 0.72 \times T_c|\frac{dH_{c2}}{dT}|_{T_c} = 13\ \text{kOe} \quad (3)$$

$H_{c2}$ (1.9 K) at various pressures are recorded by $R(H)$ measurement up to 90 kOe, (Fig. A3) where the $H_{c2}$ is denoted as an onset of the resistance drop. The evolution of $H_{c2}$ (1.9 K) inferred from experimental data is qualitatively consistent with the fitted $H_{c2}(0)$, shown in Fig. A4.

As shown in Fig. A2, obviously, the $H_{c2}(T)$ curve is far deviated from the WHH plot, indicating that the superconductivity in Re$_3$Ge$_7$ under high pressure might be a multiband superconductivity rather than a single band. The measurements at lower temperature can help to accurately estimate the contribution from different bands.

On the other hand, when both orbital and spin limiting fields are present in Re$_3$Ge$_7$, the resulting critical field will be.

$$H_{c2} = \frac{H_{c2}^{orb}(0)}{\sqrt{1+\alpha^2}} = 1.28\ T \approx 71\% \times H_{c2}^{G-L}(0) \quad (4)$$

where α is the Maki parameter,

$$\alpha = \sqrt{2}\frac{H_{c2}^{orb}(0)}{H_{c2}^p(0)} = 0.18 \quad (5)$$

The calculated $H_{c2}$ value is about 30% smaller than $H_{c2}(0)$ and very close to $H_{c2}^{orb}(0)$, suggests that the effect of spin limit is not dominant or absent, which may also indicate a spin singlet mechanism for the superconductivity. Then multiband superconductivity could be a plausible explanation. However, our experimental data do not give clear enough evidence for electron pairing mechanism in Re$_3$Ge$_7$. Further investigations are necessary.

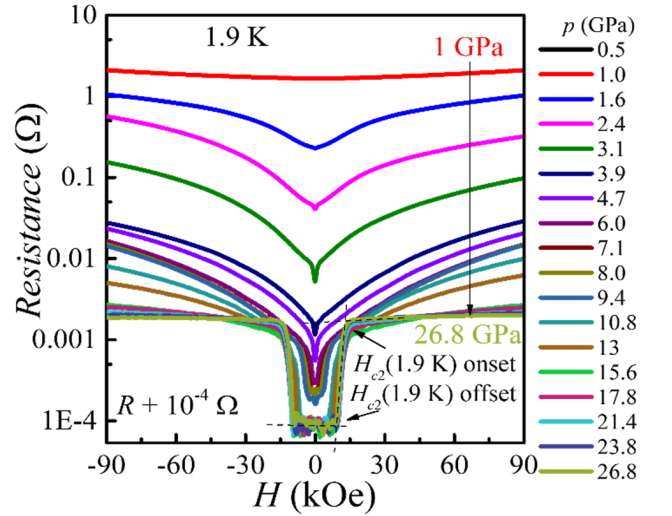

Fig. A3 Resistance as a function of external magnetic field, $R(H)$, at different pressures, measured in DAC. The magnetic field direction is parallel to $c$-axis. All $R(H)$ curves are shifted by $+10^{-4}$ Ω for a log scale plot. The criterions of $H_{c2}(1.9K)\ onset$ and $H_{c2}(1.9K)\ offset$ are the intersections of two extended lines along the $R(H)$ curves above and below the transition as shown in the figure.

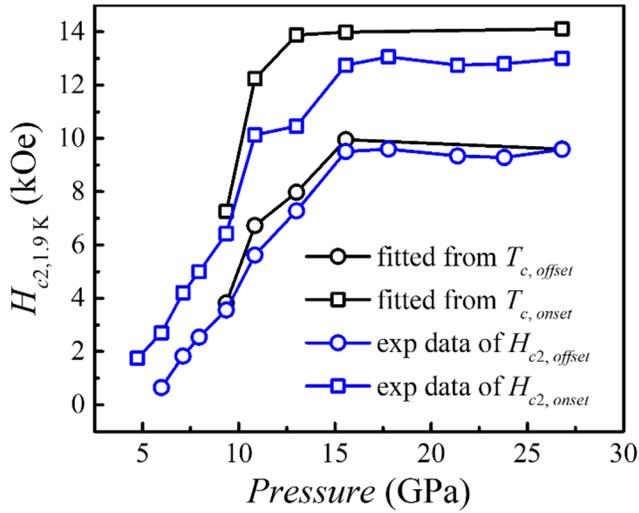

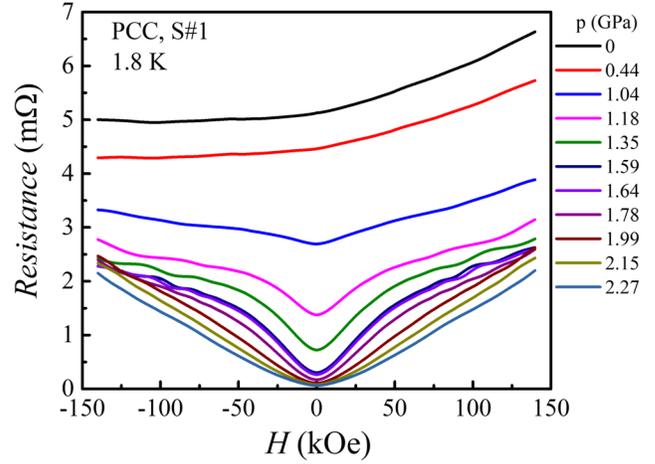

Fig. A6 Resistance as a function of external magnetic field, $R(H)$, at different pressures, measured in PCC.

Fig. A4 Summarized evolution of upper critical fields estimated by different criterions as a function of the pressure with estimated from G-L fitting at 1.9 K and from $R(H)$ data at 1.9 K. Note: the magnetic field direction is parallel to $c$-axis in all measurements.

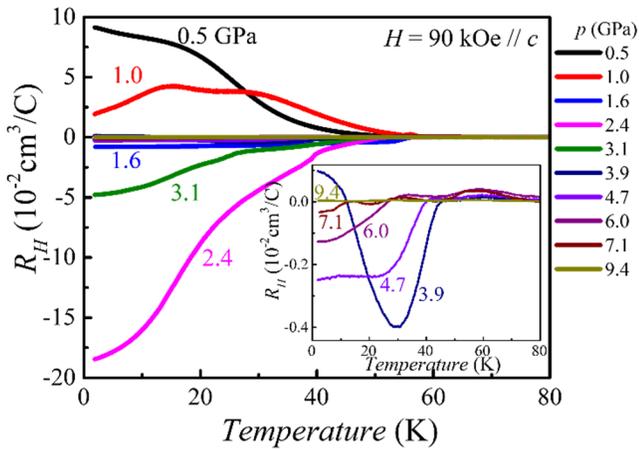

Fig. A5 temperature dependence of Hall resistance ($R_H$) at various pressures. The magnetic field is ±90 kOe and is parallel to $c$-axis. The inset shows the enlarged data from 3.9 GPa to 9.4 GPa.